# Science with Liquid Mirror Telescopes: Season 1996 of NASA Orbital Debris Observatory

Rémi A. Cabanac
cabanac@phy.ulaval.ca
Université Laval, Département de Physique, Pav. Vachon, Ste-Foy, G1K 7P4, Canada
tel. (418) 656-2131 ext 6192, fax (418) 656-240

**Abstract**: We present preliminary results of observations made, during the spring of 1996, with the 3-m liquid mirror telescope of the NASA Orbital Debris Observatory. A 20-arcmin-wide strip of sky was observed at a latitude of 32° 56' N using 10 narrow-band filters. Although the detector and the optical corrector were not optimized for astronomical observations, valuable information could be extracted down to magnitude 19 in R. From these observations we are generating a catalog of positions and AB magnitudes of objects in the strip. In this present work, we show images, maps and counts representative of the dataset.

## NASA Orbital Debris Observatory

The NASA Orbital Debris Observatory (NODO) built by NASA in Cloudcroft (New-Mexico) is equipped with a 3-m LMT. (Table I)

**TABLE I: NODO Specifications**

| | |
|---|---|
| Latitude | 32.96° N |
| Average Seeing | 1.5 - 2 arcsec |
| Mirror Diameter | 3 m |
| Focal Length | 5.18 m |
| CCD camera | 2048 x 2048 pix |
| CCD scale | 0.598 arcsec/pix (DEC)<br>0.709 arcsec/pix (RA) |
| Readout noise | 28 $e^-$ |
| Field of view | 20 arcmin |
| Integration Time | 97 sec |

NODO was not primarily designed for astronomical observations but rather for detection of satellite debris. In particular, the optical corrector at prime focus induces a small distortion on the north and south edges of the field and the CCD camera uses a fast-reading technique optimized for moving objects generating a readout noise of 28 $e^-$. This high readout noise introduces severe limits on the lower value of the detectable sky. For filters bluer than 650 nm, the frames are not sky limited but rather noise limited. Another limit of the data comes from the fact that the observed objects don't follow linear trajectories through the CCD field, but follow circles around the north pole. At NODO latitude, the differential drift between the center and the edge of the frames is 3.3 arcsec (Stone, Monet et al. 1996). CCDs cannot correct for this effect unless they have a parallel clock for each column. Another way to correct for the drift would be to add a wedge-lens in the optical path. The NODO was not equipped with either device. However, in our case, the drift is rather negligible for an average seeing of 2 arcsec, except for the very edge of the frames where the stars have triangular PSF.

## Qualitative description of 1996 dataset

During the spring of 1996, Hickson and Mulrooney (1997) undertook a program of astronomical observations using a set of narrow-band interference filters designed for the Liquid Mirror Telescope Survey (LIMITS) project of Laval-UBC 2.7-m LMT (Hickson, Borra et al. 1994; Cabanac and Borra 1995). They observed for 34 nights with 11 filters at 450, 500, 550, 600, 650, 700, 750, 800, 850, 900, 950 nm. Each night provided a long strip of sky made of 280 frames (2k x 2k). Figure 1 is a description of the dataset. Unfortunately a disk crash erased the logbook and we lost the exact dates of observation for each night. Hence, we don't know the exact phases of the Moon and we cannot correct for gradients in our frames. Few nights were photometric during the New Moon dark periods. The sky of most of the clear nights suffered from North-South gradients, i.e. perpendicular to scanning. The amplitude of these gradients varied from 1% to 20% of the sky intensity. The typical area covered during one night is 20 deg$^2$ from the galactic pole to the galactic plane (b<25°).

**Figure 1: Season 1996 of NODO**

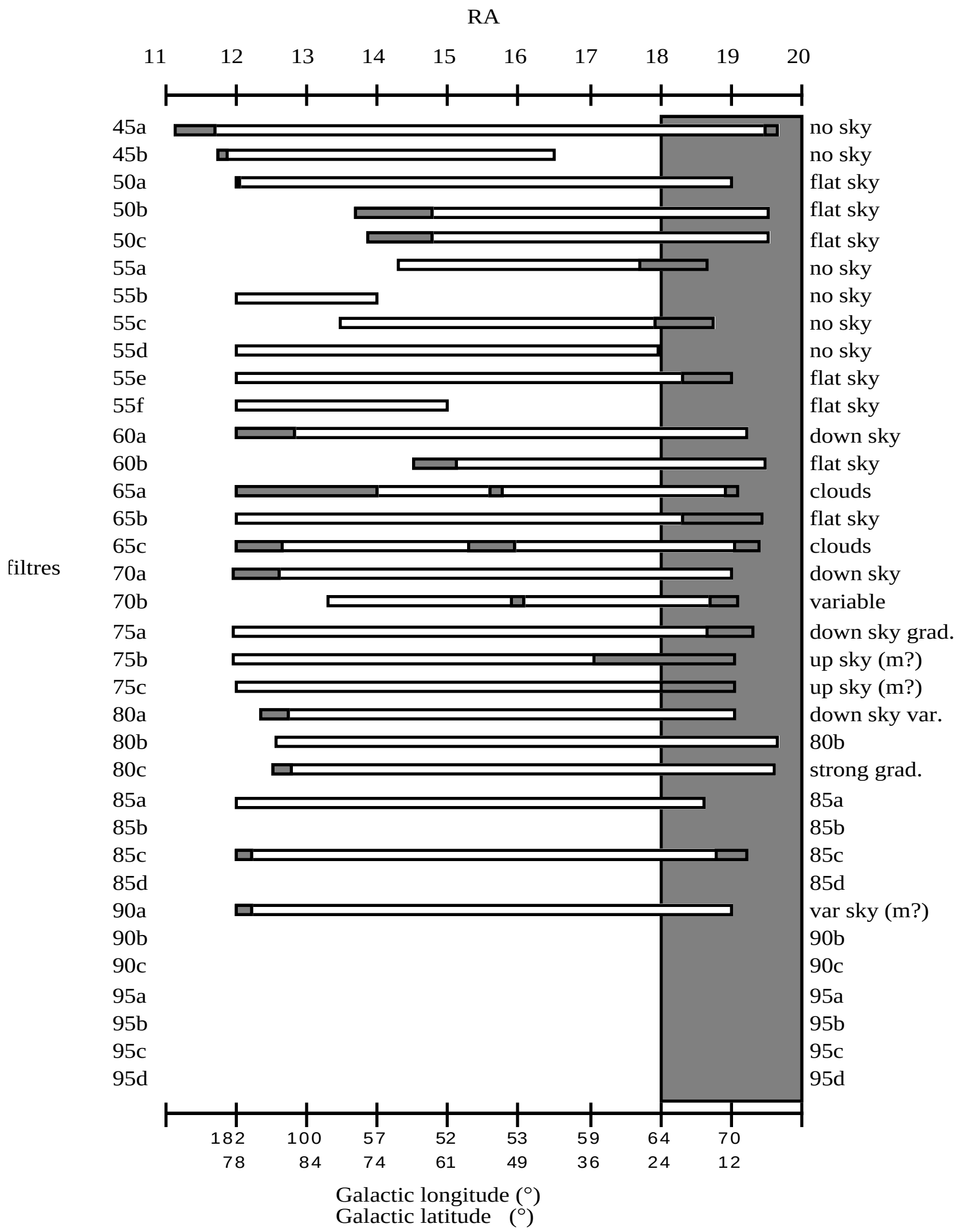

**Figure 2**

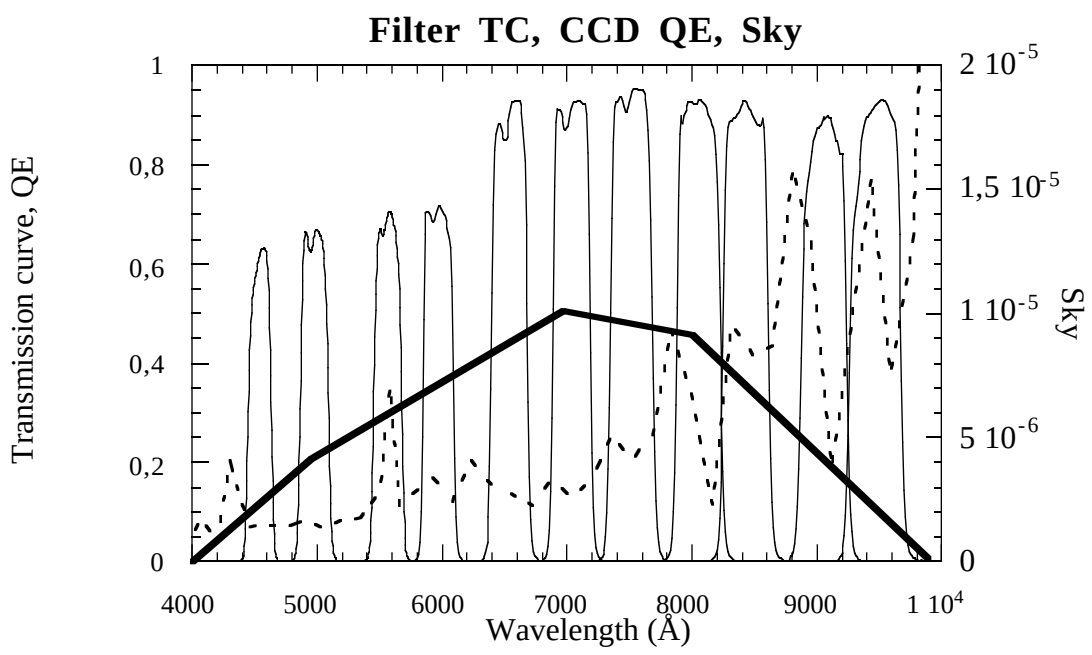

Figure 2 gives the typical quantum efficiency (QE) for a thick 2kx2k, the sky brightness from Kitt Peak, and the transmission curves (TC) of the filters. It can be seen that the QE was poor (<10%), combined with the low TC of the filters under 650 nm (TC<60% and bandpass <150Å) in the blue part of the spectrum. Since sky brightness increased exponentially after 900 nm, we did not expect to detect faint objects with filter 45 or filter 95.

In order to visualize the quality of our frames, we compared fields randomly chosen in our dataset with the corresponding fields from the digitized sky survey (DSS) and we searched for any difference that might be visible in deepness, scattered light around bright objects, visual aspect, etc. Figure gives an example of an image taken by the NODO LMT without filter, i.e. with a peak of detection near 750 nm 3(a), and the same field extracted from the red plate of the DSS 3(b). The latter was extracted from the first generation DSS with a scale of 1.7 arcsec/pix. One can see that these images are remarkably similar in deepness except for the better sampling and seeing of the NODO frame.

In our data blue frames are shallower than red ones for the reasons mentioned above (poor QE of CCD and low TC of blue filters), they did not go as deep as the POSS O plates. Yet, from this picture one may claim that NODO frames go as deep as POSS in red filters (>700 nm). Scattered light is not dominant although some frames showed large wings around the bright stars. Girard and Borra (1997) studied this scattered light effect. Its origin is now understood and we control it with a good levelling of the LMT and thin layers of mercury (Girard 1996). In the following the data will be quantitatively analyzed e.g. in number counts and magnitude counts in red. To achieve such a goal it was necessary to reduce first the raw data.

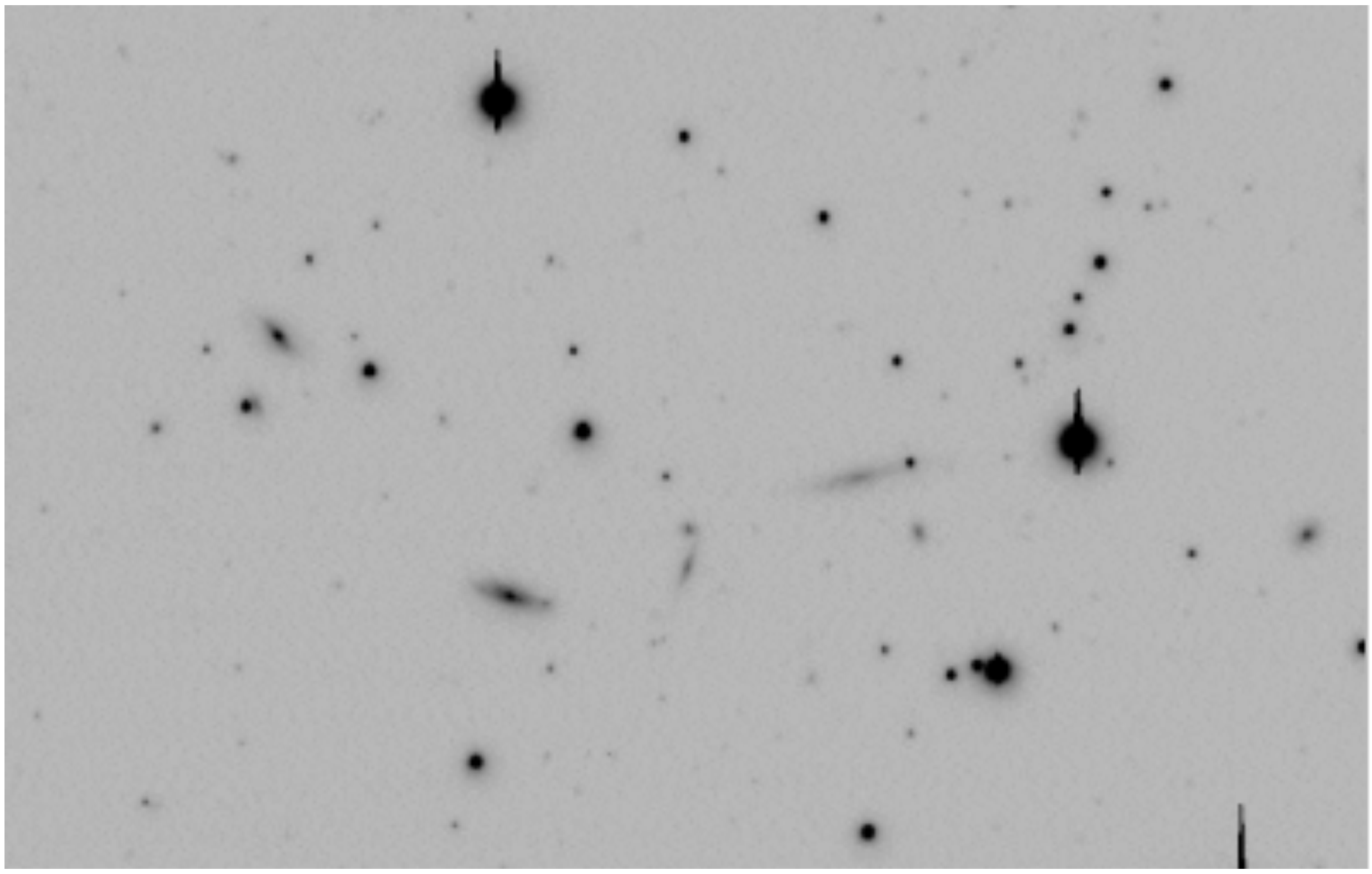

**Figure 3(a): NODO frame 5'x7', no filter, 97s**

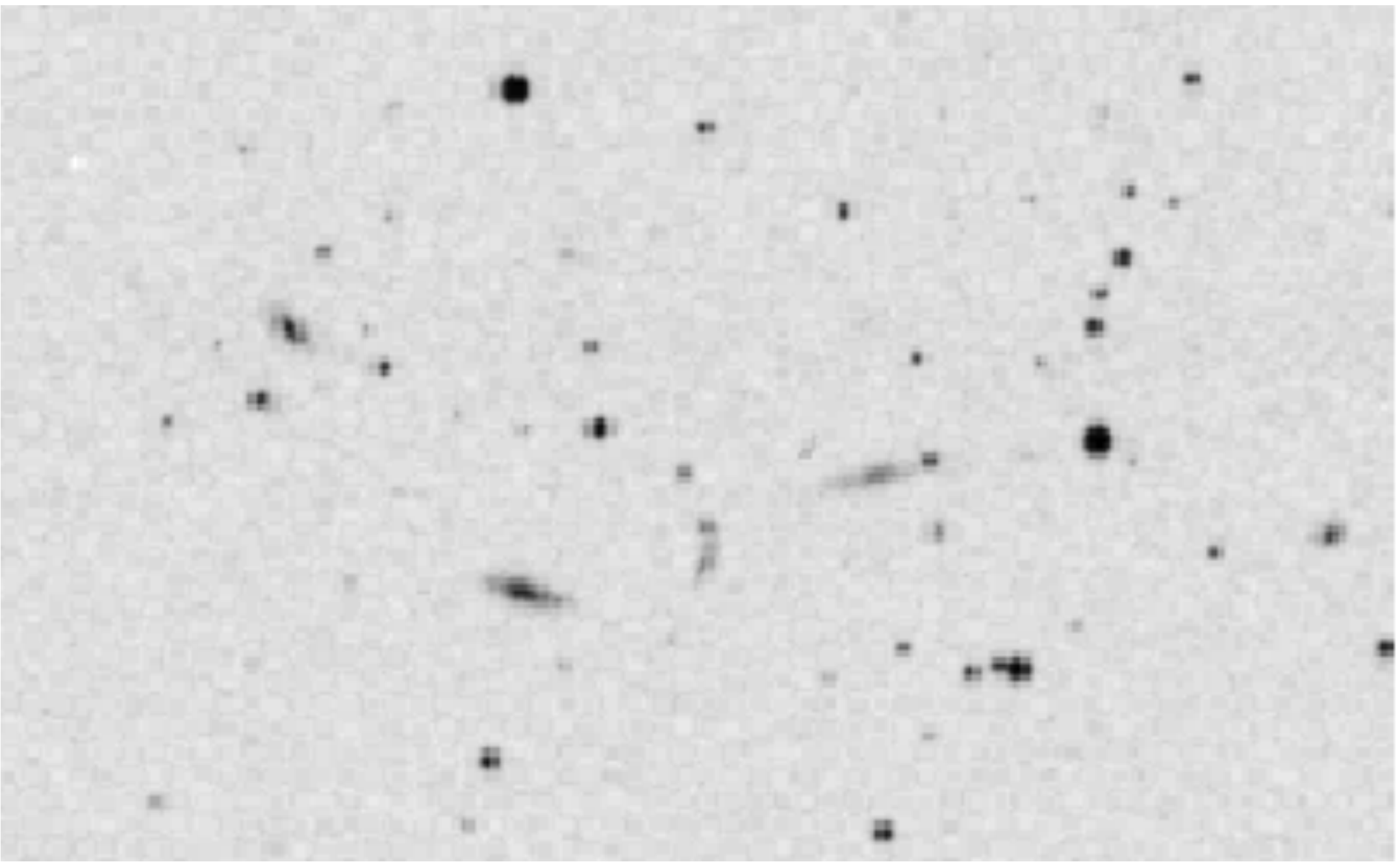

**Figure3(b): POSS E plate from DSS, 5'x7', 45min**

**Data reduction**

The entire dataset was roughly 60 Tb. In order to reduce this huge amount of information we wrote a series of C routines optimized for the format of the images. Each night was cut into about 280-300 2kx2k frames with an overlap of one arcmin between them. The set of automated C routines reduced the data according to the following steps:

1. Remove bad pixels and columns and average the gap from the nearest pixels and columns.
2. Convolve image with 3x3 gaussian kernel (FWHM=2.35pix=1.4 arcsec).
   This reduces sky noise when available and increases the number of bits in sky histogram.
3. Detect all pixels above 2 $\sigma$ of sky mosaic (32x32).
   Due to a strong gradient in some of the nights we had to adjust the size of mosaic elements in order to have enough pixels to get a good sky value and $\sigma$ in one hand, and to keep sky gradient through each mosaic element under 1%.
4. Identify groups bigger then 9 contiguous pixels and tag them as objects.
   This was our primary threshold to segregate true objects and false detections. This loose threshold allowed 50% to 80% of detections to be false objects because we didn't want to lose actual objects.
5. Measure centroid, ellipticity, angle, Kron radii around every objects.
6. Measure growth curve from 2 pix to 10 pix, by step of one pixel.
7. We also measure the flux inside Kron radius.

## Photometric calibration

To make a zero-point calibration of the nights we used the spectra of two blue stars which were available and calibrated in AB magnitudes at http://www.eso.org/spect-phot-standards/optuvstandards.html (Oke 1990). Theoretically, one should use several stars through the night to be sure that absorption does not vary with time. However, for the first projects, we did not require strong absolute photometry. A set of secondary calibrators was obtained from the same data by Paul Hickson during Spring 1997 at Kitt Peak. This should significantly improve zero-point calibrations.

## Astrometric calibration

Once primary reduction and photometric measurements were done, the next important step was to achieve a fine astrometry over the whole night. Coordinates of objects are the central criterion for identification and merging of objects from one night to another. We could have used the magnitude, but then would have eliminated all sources that vary intrinsically.

The astrometry was calibrated in two steps, first we used analytic transformations from instrumental coordinates to J2000 equatorial coordinates. This allowed to correct for precession, nutation, and aberration. Because LMTs are transient instruments, they are particularly sensitive to aberration (up to 20 arcsec from the beginning to the middle of the night). For the second step we wrote a fine-tuning program, using the USNO astrometric survey available at http://cadcwww.dao.nrc.ca/usno/. We correct the coordinates frame by frame comparing all matched stars (usually over 50/frame, up to 500). The final astrometry was good within 0.5 arcsec.

## Results: Star counts, colors and magnitudes

Preliminary results presented here were extracted from the catalog. Figure 4 is a map of the strip with 30000 objects detected in all three colors; 500, 700 and 850 nm (one night per filter).

**Figure 4**

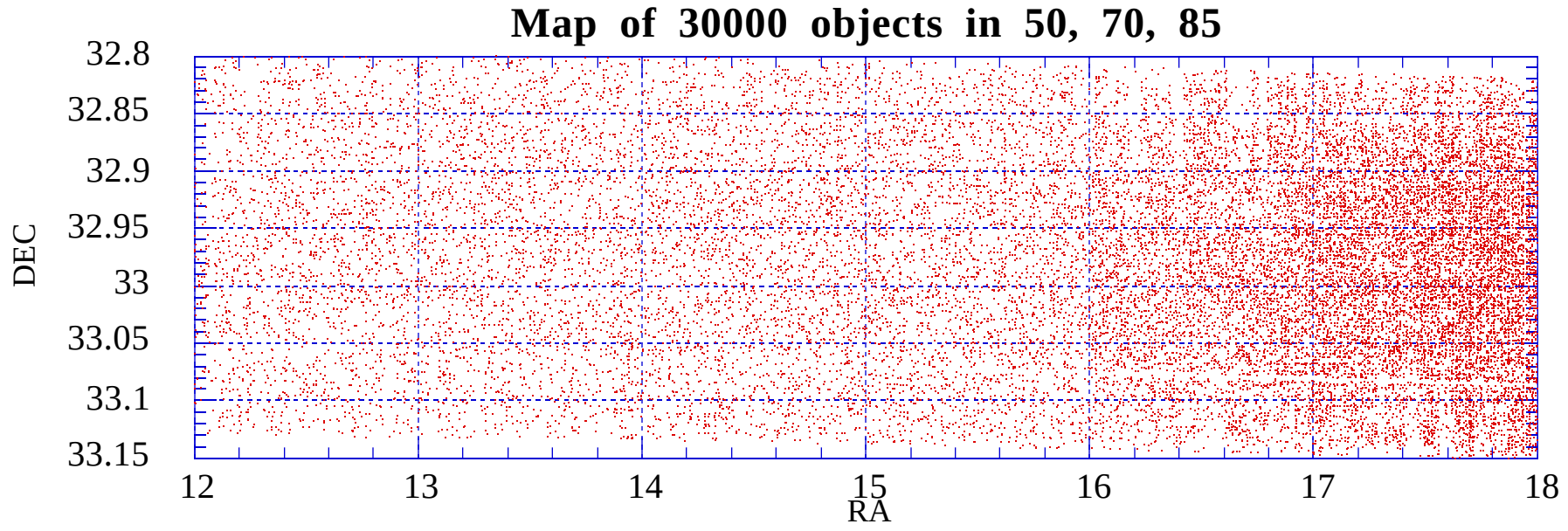

We see that the number of objects rose fast as we approached the galactic plane. We also see that detection of objects was not homogeneous near the galactic plane (for RA>16). It is not clear, yet, whether this resulted from bad weather, from confusion because of overcrowding objects, or simply from actual structure in the galactic plane compressed in RA.

Figure 5 is a summary of pieces of information available from our dataset. Figure 5(a) is a color-magnitude diagram of the objects of Figure 4. Since we did not know the distances of our objects and since we did not segregate extended objects from stars in Figure 5(a), the picture is noisy. Nevertheless, this color diagram is similar to what can be extracted from the APM survey (see Gilmore, King et al. 1990). Figure5 (b) plots number counts against RA. As expected from the qualitative comparison of frames, one can see that red filters go as deep as the USNO catalog of the POSS. We must keep in mind that the USNO astrometric catalog is not meant to give precise spectrophotometry, and that it keeps only objects seen in both E and O plates down to blue=21 and red=20. Therefore, Figure 5(b) should not be considered as a definite test. Figure 5(c) is a magnitude count diagram using same data as 5(b). Since blue nights are shallower we cannot expect data obtained with blue filters (50) to overlapp the USNO magnitude counts, but we follow the counts nicely with red filters (75-85). The slight difference in slope between the upper lines was not significant given the uncertainty in USNO photometry. Figure 5(d) is the color distribution of objects in a given range of magnitudes and galactic latitudes. This is similar to plots of Lasker, Garnavich et al. (1987) who used the Guide Star Photometric Catalog. We plan to compare this distribution with Galaxy models of Bahcall (1986); Ibata (1997).

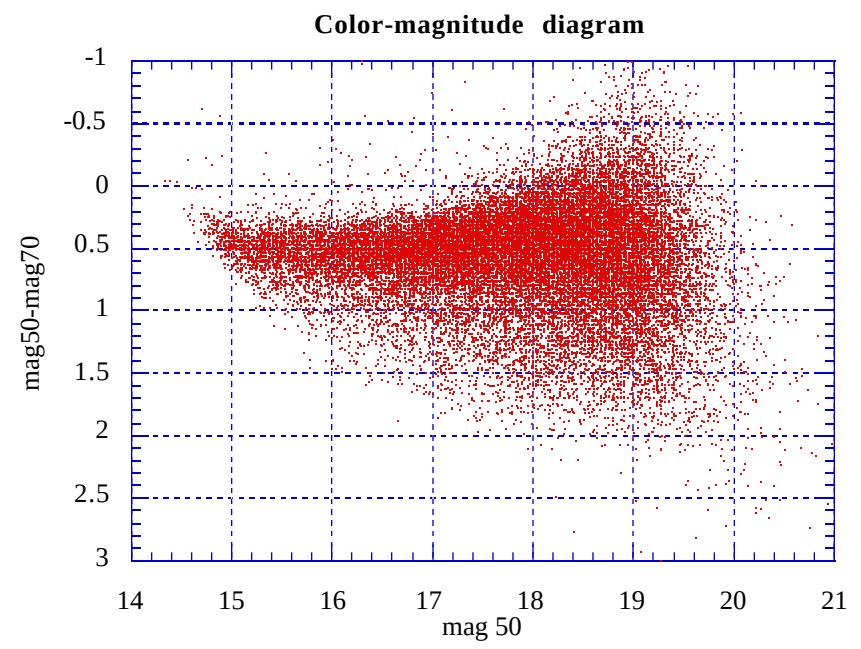

**Figure 5(a)**

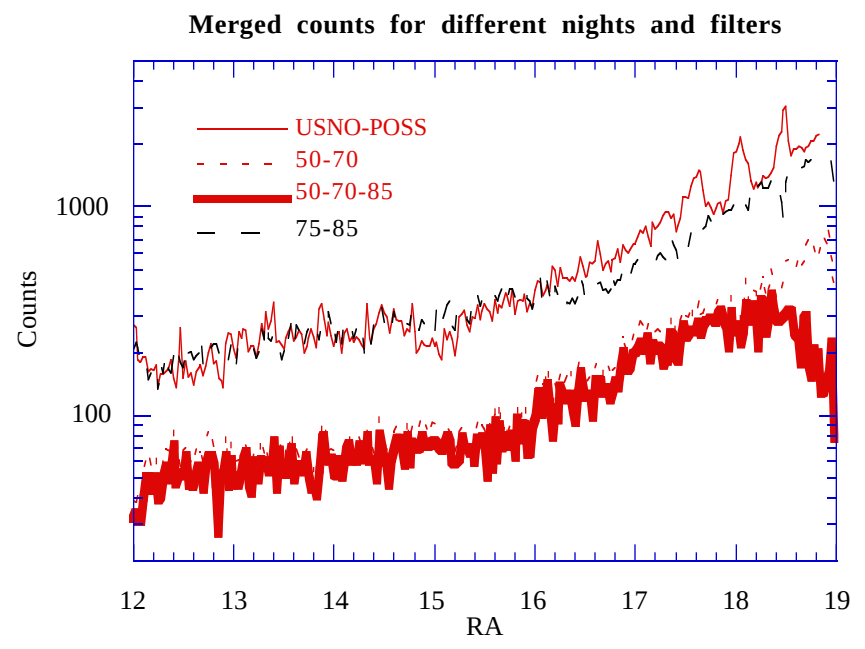

**Figure 5(b)**

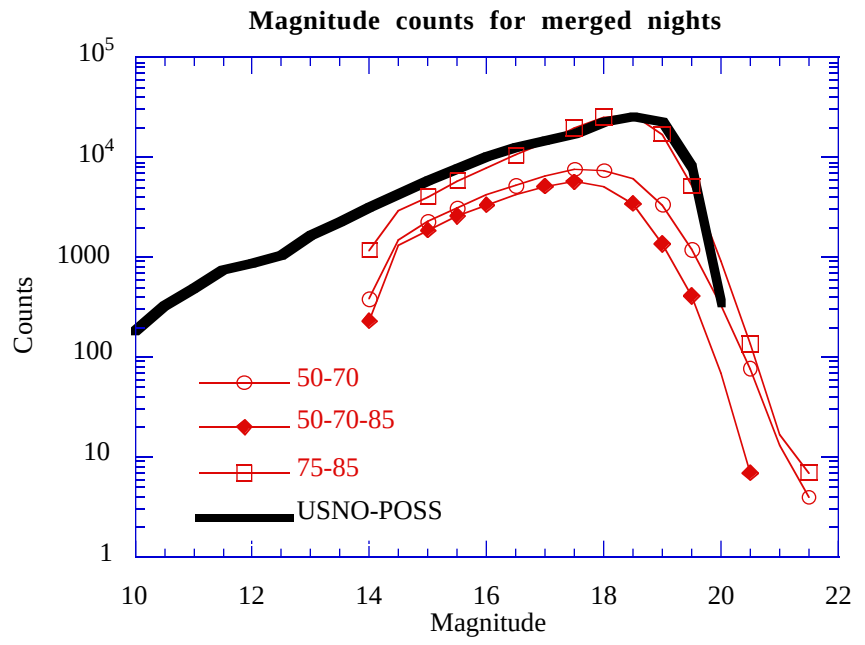

**Figure 5(c)**

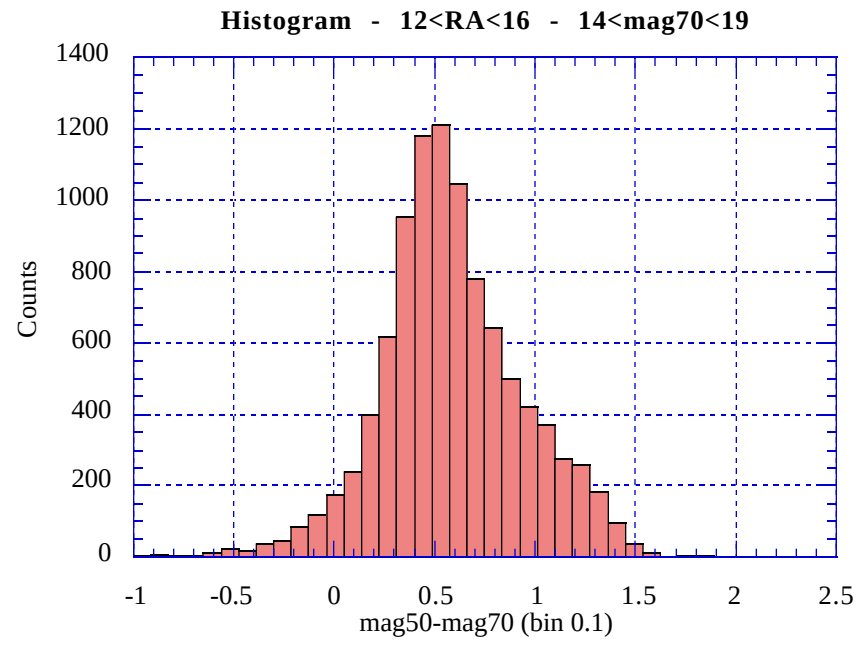

**Figure 5(d)**

We are currently finishing to reduce the dataset of season 1996. We think that this dataset already demonstrates the usefulness of LMTs in the field of surveys.

## Acknowledgments

The author would like to thank Hickson and Mulrooney from giving their dataset prior to publication.

This work has been made possible by the use of the USNO-A astrometric Catalog using POSS I (Monet 1997) and the Digitized Sky Survey produced at the Space Telescope Institute under U.S. Gov. Grant NAG W-2166. This work was supported by NSERC grants to Ermanno F. Borra, P. Hickson, and NSERC fellowship to RAC.